\documentclass[pre,twocolumn,
showpacs, eqsecnum
]{revtex4}
\newcommand{\be}{\begin{equation}}
\newcommand{\ee}{\end{equation}} 
\newcommand{\bea}{\begin{eqnarray}} 
\usepackage{graphicx}
\newcommand{\eea}{\end{eqnarray}}
\usepackage{epsfig}
\usepackage{bm}

\begin{document}
\title{Instanton Theory of Burgers Shocks and Intermittency}
\author{L. Moriconi}
\affiliation{Instituto de F\'\i sica, Universidade Federal do Rio de Janeiro, \\
C.P. 68528, 21945-970, Rio de Janeiro, RJ, Brazil}
\begin{abstract}
A lagrangian approach to Burgers turbulence is carried out along the lines of the field 
theoretical Martin-Siggia-Rose formalism of stochastic hydrodynamics. We derive, from
an analysis based on the hypothesis of unbroken galilean invariance, the asymptotic
form of the probability distribution function of negative velocity-differences. 
The origin of Burgers intermittency is found to rely on the dynamical 
coupling between shocks, identified to instantons, and non-coherent background 
fluctuations, which, then, cannot be discarded in a consistent statistical description 
of the flow.

\end{abstract}
\maketitle
\section{Introduction}
The long established Burgers model of compressible one-dimensional flow \cite{burgers}
provides an interesting testing ground for the performance of numerical and analytical
strategies in turbulence research. Despite its own peculiar phenomenology,
as evidenced by the complete failure of approximations based on the ``K41" scaling 
theory \cite{frisch}, there are important conceptual analogies between the Burgers model 
and usual three-dimensional turbulence (see Ref. \cite{bec1} for a comprehensive 
review).

We note, in passing, that the Burgers model is more than just a ``mathematical toy"; in 
its multidimensional version the Burgers equation plays an important role in several 
realistic problems, like nonlinear acoustics \cite{gurba-etal}, cosmology 
\cite{zeldo,gurba-saiche}, critical interface growth \cite{kpz}, and traffic flow 
dynamics \cite{chowd-etal}.

A great deal of attention has been focused on the problem of non-gaussian fluctuations observed in
the high Reynolds number regime of Burgers dynamics -- the intermittency phenomenon, 
for short. As it is verified through numerical simulations \cite{chekhlo-yakhot,gotoh-kraichnan}, 
velocity differences
\be
z = u(x+\zeta,t) - u(x-\zeta,t) \label{vel-diff}
\ee 
are found to be very intermittent at small scales ($\zeta$ much lesser than the integral scale $L$). 
The probability distribution function (pdf) of the right tail of $z$, which decays faster than gaussian, has been 
analytically obtained in a number of different ways \cite{gotoh-kraichnan,gura-migdal,poly,weinanE-etal}. The left 
tail, on the other hand, which is related to Burgers shocks, is found to have a power law profile 
$\rho(z) \sim 1/|z|^{\alpha}$, with no sharp consensus on the value taken by the exponent $\alpha$.

A Fokker-Planck approach to the computation of velocity-difference pdfs, with closure given by
an operator product expansion treatment of the dissipative anomaly was put forward by Polyakov \cite{poly}. 
This method provides a fine description of the pdf's right tail, and yields a power law form for 
the left tail with $5/2 \leq \alpha \leq 3$ \cite{boldy}. Extensive numerical simulations performed by Gotoh 
and Kraichnan \cite{gotoh-kraichnan} indicate that $\alpha = 3$. At variance with such findings, an analytical study based 
on the velocity field profiles in space-time neighborhoods of shocks, the so-called preshock events, gives 
$\alpha = 7/2$ \cite{weinanE-etal,weinanE-eijn}, a result confirmed by alternative lagrangian simulations of the Burgers 
equation \cite{bec2,gura}. 

In an attempt to conceal these apparently contradictory conclusions, Boldyrev et al. have suggested that the left 
tail exponent is not universal, departing from $\alpha=3$ if flow realizations fail to satisfy a strong form of galilean 
invariance \cite{boldy-etal}, which holds, by definition, if usual galilean invariance is observed in the bulk, regardless 
the boundary conditions at infinity. In rephrased form, the whole point of Ref. \cite{boldy-etal} is that finite-size 
effects which break strong galilean invariance would lock larger fluctuations of shock jumps and negative velocity 
derivatives, reducing intermittency. In this paper, we find support to the conjecture that the left tail exponent is 
$\alpha =3$ when the strong form of galilean invariance is fulfilled.

This work is organized as follows. In section II, we introduce Burgers intermittency as a phenomenon related, in 
the inviscid limit, to shock amplitude fluctuations. The great convenience of a lagrangian description of the flow is then 
pointed out. In section III, we discuss the Martin-Siggia-Rose (MSR) formulation of stochastic hydrodynamics \cite{msr,phythian} 
within the lagrangian perspective. In section IV, Burgers shocks will be given as instantons \cite{gura-migdal, falko_etal}, 
and background fluctuations around them will be taken into account in the computation of the asymptotic behaviour of the pdf 
of negative velocity differences. In section V, we summarize and discuss our results. 

In appendices A and B we provide technical details on some of the material discussed in section IV.

\section{Phenomenological Considerations}

The Burgers model describes the dynamics of a one-dimensional velocity field,
$u=u(x,t)$, ruled by the evolution equation
\be
\partial_t u + u \partial_x u = \nu \partial^2_x u + f \ , \  \label{burgers-eq}
\ee
where $\nu$ is the kinematical viscosity and $f = f(x,t)$ is the external force which sustains the flow and introduces the integral length scale $L$.
There is no pressure term in the above equation, and no imposition of incompressibility as well. 

Let $u_0(x) \equiv u(x,0)$ be the velocity field at initial time, supposed to be a $C^1$ function defined on $- \infty < x < \infty$.  The Cauchy problem is exactly solvable for the Burgers equation \cite{hopf, cole}. In the forceless case, the velocity field is, at time $t>0$,
\bea
u(x,t) &=& - 2 \nu \partial_x \ln  \left \{ \int_{- \infty}^\infty dy \exp  \left [ - \frac{(x - y)^2}{4 \nu t} \right. \right. \nonumber \\
&-&\left. \left.   \frac{1}{2 \nu} \int_0^{y} dx' u_0(x') \right ] \right \} \ . \ \label{u-sol}
\eea
As it is well-known, (\ref{u-sol}) leads, in the vanishing viscosity limit, to discontinuous shocks (i.e., the velocity field becomes piecewise 
$C^1$), which can be interpreted as ``sinks" of fluid particles. External forcing does not spoil the process of shock creation, even though it can affect the statistics of shock amplitudes.

Stable and unstable regions of the flow are distinguished, essentially, by the sign of the spatial velocity derivative. Neglecting higher order corrections, consider the expansion $u(x,t) = \sigma_0(t) + \sigma_1(t) x$ in the neighborhood of an arbitrary point. Eq. (\ref{burgers-eq}) gives, for
$f(x,t)=0$,
\bea
&&\dot \sigma_0 + \sigma_0 \sigma_1 = 0 \ , \ \nonumber \\
&&\dot \sigma_1 + \sigma_1^2 = 0 \ , \
\eea
leading to
\be
\sigma_0(t) = \sigma_0(0) \exp \left [ - \int_0^t dt' \sigma_1(t') \right ] \ , \
\ee
where
\be
\sigma_1(t) = \frac{\sigma_1(0)}{1+\sigma_1(0) t} \ . \ \label{u-grad}
\ee
Therefore, if $\sigma_1(0) = \partial_x u |_{t=0}$ is positive, we expect, from (\ref{u-grad}), that $\partial_x u$ will decay. On the other hand,
if $\sigma_1(0)$ is negative, then $|\partial_x u|$ will increase in time, implying flow instability. This is the mechanism for the generation of large negative velocity derivatives in Burgers turbulence, which in a time scale of order $1/|\partial_x u|$ are transformed into long lived shocks.

We are interested to study the statistics of negative velocity differences $z$ for $\nu \rightarrow 0$ and $\zeta / L \ll 1$. Under these circumstances, shock jumps provide the main contribution to the strong fluctuations of $z$. A time series of $z$ would exhibit intermittent negative spikes associated with the passage of shocks, separated by much weaker signals due to smooth velocity configurations. 

Suppose, now, that fluctuations of $z$ are alternatively measured from the subtraction of velocity fields defined at points $x(t)+\zeta$ and $x(t)-\zeta$, where $x(t)$ is the position of a fluid element that moves with the flow. We expect to have, in such a lagrangian framework, the same asymptotic power law form for the left tail pdf of $z$. The central point in this correspondence is that fluid particles typically spend a finite fraction of their times at shock discontinuities. Once a fluid particle is dragged into a shock discontinuity, it remains there until the shock collapses, or it is absorbed by another one.

Negative spikes in the eulerian time series of $z$ are replaced, in the lagrangian reference frame, by smooth fluctuations of shock amplitudes, which 
last for much longer times. In order to compute statistical properties of the eulerian negative spikes, one would have to find out how shocks match to each other in solutions of the Burgers equation. Within the lagrangian framework, on the other hand, it suffices to describe fluctuations around isolated shocks -- a much simpler task that points out, for our purposes, the advantage of lagrangian methods over eulerian ones.
 
\section{Path-Integral Framework}
In the stochastic hydrodynamics approach to Burgers turbulence, the external forcing in (\ref{burgers-eq}) is taken to be a large scale gaussian random 
field, with zero mean and correlator
\be
\langle f(x,t) f(x',t') \rangle  = D(|x-x'|) \delta(t-t') \ , \
\ee
where we take
\be
D(|x-x'|) = D_0 \exp(-|x-x'|^2/L^2) \ . \ 
\ee

The conditional probability density functional to have velocity configuration $u_0(x)$ at time $t=0$ 
if $u_{-T}(x)$ is the velocity at time $t=-T$, can be written as the path integral \cite{msr,phythian}
\be
Z = {\cal{N}} \int D \hat u D u \exp(iS) \ , \  \label{msr}
\ee
where ${\cal{N}}$ is an unimportant normalization factor (which will be usually supressed from notation) and
\bea
&&S = S[\hat u, u] \equiv \int_{-T}^0 dt \int dx \hat u [ \partial_t u+ u \partial_x u - \nu \partial^2_x u ]
\nonumber \\
&& + \frac{i}{2} \int_{- T}^0 dt \int dx dx' \hat u(x,t) \hat u(x',t) D(|x-x'|) \label{msr-action}
\eea
is the so-called MSR action. Expressions (\ref{msr}) and (\ref{msr-action}) are subject 
to the boundary conditions
\bea
u_0(x) &=& u(x,0) \nonumber \\
u_{-T}(x) &=& u(x,-T) \ . \
\eea

In order to pave the way for a lagrangian description of the flow, let us consider a general reference frame $R'$ which moves
with velocity $\phi(t)$ relative to the original (inertial) ``laboratory" frame $R$. The position and velocity in $R'$ are
\bea
&&x' = x -\int_{-T}^t dt' \phi(t') \ , \ \label{g-gt1} \\
&&u_\phi (x',t) = u(x,t) - \phi(t) \ . \ \label{g-gt2}
\eea 
In the moving frame, the velocity at the origin ($x'=0$) is defined as
\be
u_\phi(t) \equiv u_\phi(0,t) = u\left (\int_{-T}^t dt' \phi(t'),t \right ) - \phi(t)
\ee
For a given field $u=u(x,t)$, there is a unique time-dependent function $\phi(t)$ which solves $u_\phi(0,t)=0$. It is clear that $\phi(t)$ is in this case the velocity of a locally comoving reference frame -- that's how lagrangian coordinates come into play. We introduce, correspondingly, the Faddeev-Popov determinant \cite{faddeev-popov}, $\Delta[u(0,t)]$, by means of
\be
\Delta^{-1}[u(0,t)] \equiv \int D \phi \delta [ u_\phi(t) ] \ . \ \label{fp}
\ee
Note that $\Delta[u(0,t)]$ is invariant under the generalized galilean tranformations given by (\ref{g-gt1}) and (\ref{g-gt2}). In fact,
\bea
&&\Delta^{-1}[u_{\phi_0}(t)] \equiv \int D \phi \delta [ u_{\phi+\phi_0} (t)]  \nonumber \\
&&= \int D \phi \delta [ u_\phi(t) ] =  \Delta^{-1} [u(0,t)] \ . \
\eea
Relation (\ref{fp}) yields
\be
\Delta[u(0,t)] \int D \phi \delta [ u_\phi(t) ] = 1 \ . \ \label{fp2}
\ee
Inserting (\ref{fp2}) into the integrand of (\ref{msr}) and exchanging the order of integrations, we get
\be
Z = \int D \phi \int D \hat u D u \Delta[u(0,t)] \delta [ u_\phi(t) ] \exp (iS) \ . \ \label{msr2}
\ee

Generalized galilean transformations can be used to replace the Dirac's delta functional in (\ref{msr2}) by
$\delta [ u_{\phi=0}(t) ] = \delta [ u(0,t)]$. To accomplish that, we first substitute, in the MSR action of (\ref{msr2}), 
the integration fields $u(x,t)$ and $\hat u(x,t)$ by galilean transformed ones, through
\bea
&& u(x,t)= u_\phi ( x' ,t) + \phi(t) \ , \ \label{g-gtb1} \\
&&\hat u (x,t) = \hat u_\phi ( x',t) \ . \ \label{g-gtb2}
\eea 
We find
\be
S= S_\phi + \int_{- T}^0 dt \int dx \hat u_\phi \frac{d \phi }{dt} \ , \ \label{s_phi}
\ee
where
\bea
&&S \equiv S[\hat u(x,t),u(x,t)] \ , \  \\
&&S_\phi \equiv S[\hat u_\phi (x,t),u_\phi (x,t)] \ . \
\eea
The additional term on the RHS of (\ref{s_phi}) takes account of the non-inertial force due to the acceleration $\dot \phi$ of the reference
frame $R'$.

The jacobian associated with the above transformations is unity, as can be verified from the matrix elements
\bea
&&\hat O(x_1,x_2 | t_1,t_2) \equiv \frac{ \delta \hat u(x_1,t_1)}{\delta \hat u_\phi(x_2,t_2)} = \frac{ \delta u(x_1,t_1)}{\delta u_\phi(x_2,t_2)} \nonumber \\
&&= \delta \left ( x_1-x_2+\int_{- T}^{t_1} dt' \phi(t') \right ) \delta(t_1-t_2) \ . \ \label{matrix-elems}
\eea
The operator which has the matrix elements (\ref{matrix-elems}) can be written in any reasonable functional space of space-time 
dependent functions as
\be
\hat O = \exp \left ( \int_{-T}^{t} dt' \phi(t') \frac{\partial}{\partial_x} \right) \ . \
\ee
The eigenstates of $\hat O$ are the momentum wavefunctions $ \exp(ipx) $. 
Using a parity-preserving discretization of the Fourier space, the jacobian turns out to be 
\be
\det [ \hat O ] = \prod_{p} \exp \left ( i p \int_{- T}^{t} dt' \phi(t')  \right) = 1 \ . \
\ee

The Faddeev-Popov determinant $\Delta[u(0,t)]$ is also unity. In fact, consider the velocity field which has been ``gauge fixed", with the help of a generalized galilean transformation, to $u(0,t)=0$. We have, then, to substitute the functional Taylor expansion of $u_\phi(t)$ up to first order in $\phi(t)$ in (\ref{fp}). Defining $g(t) = \partial_x u(x,t) |_{x=0}$, we get
\be
u_\phi(t) = g(t) \int_{-T}^t dt' \phi(t') - \phi(t) + O[\phi^2(t)]
\ee
so that
\bea
&&\Delta^{-1}[u(0,t)] \equiv \int D \phi \delta \left [ g(t) \int_{-T}^t dt' \phi(t') - \phi(t) \right ] \nonumber \\
&&= \left | \det[\delta(t-t')- \Theta(t-t') g(t) ] \right |^{-1} \ . \ \label{fp3}
\eea
Using, now, the identity
\be
\det[X] = \exp[{\hbox{Tr}}(\ln X)] \ , \
\ee
we find, with $A(t,t') \equiv \Theta(t-t') g(t)$,
\be
\Delta[u(0,t)] = \exp \left [ \sum_{n=1}^\infty \frac{(-1)^{n+1}}{n}{\hbox{Tr}}(A^n) \right ] = 1\ , \
\ee
since
\bea
&&{\hbox{Tr}}(A^n) = \int dt_1 dt_2 \ ... \ dt_n g(t_1)g(t_2) \ ... \ g(t_n) \nonumber \\
&& \times \Theta(t_1-t_2) \Theta(t_2-t_3) \ ... \ \Theta(t_n - t_1) = 0 \ . \
\eea
Above, we have used $\Theta(0) = 0$, which is the right prescription for the Heaviside function 
when the underlying stochastic differential equation is defined in terms of an It\^o discretized time 
evolution \cite{cardy}. Collecting all of the above results, we rewrite (\ref{msr2}) as
\bea
&&Z = \int D \phi \int D \hat u_\phi D u_\phi \delta [ u_\phi(t) ] \nonumber \\
&& \times \exp \left ( iS_\phi+i\int_{- T}^0 dt \frac{d \phi }{dt} \int dx \hat u_\phi  \right ) \ . \ \label{msr3}
\eea
However, since $\hat u_\phi(x,t)$ and $u_\phi(x,t)$ are integration fields, (\ref{msr3}) becomes
\bea
&&Z = \int D \phi \int D \hat u D u \delta [ u(0,t) ] \nonumber \\
&& \times \exp \left ( iS +i\int_{- T}^0 dt \frac{d \phi }{dt} \int dx \hat u  \right ) \ , \ \label{msr4}
\eea
or, equivalently, integrating over $\phi(t)$,
\be
Z = \int D \hat u D u \delta[ \int dx \hat u(x,t)] \delta [ u(0,t) ] \exp (i S)  \ . \ \label{msr-phi-psi}
\ee

In view of (\ref{msr4}), we will assume, in all our subsequent considerations, that $u(0,t)=0$. In other words,
we have moved for good to the locally comoving reference frame. As fluid elements happen to stick (and spend a finite
fraction of their times) in shock discontinuities, the latter will be frequently hosted at the origin of the locally 
comoving reference frame. 

We stress, at this point, that the hypothesis of strong galilean invariance \cite{boldy-etal} is a fundamental ingredient 
here, since no role is given to the velocity boundary conditions at infinity, in the lagrangian formulation 
put forward in Eq. (\ref{msr4}).

\section{Shocks and Intermittency}

The response functional (\ref{msr-phi-psi}) can be decomposed as
\be
Z = Z_s + Z_{\bar s} \ , \
\ee
where 
$Z_s$ and $Z_{\bar s}$ refer, respectively, to the cases where shocks and smooth field configurations are found at $x=0$, $t=0$.
Recalling the discussion of section II, it is clear that the statistical properties of large negative velocity-differences are all 
encoded in $Z_s$. 

According to the probabilistic interpretation of $Z$, we note that $Z_s$ is not normalized to unit. Instead, $Z_s$ 
is normalized to the ``intermittency factor" $\gamma$, where $0 < \gamma < 1$ is the fraction of time a shock is found at the origin 
of the locally comoving reference frame. 

Let $u_s(x)$ be a shock configuration, with velocity discontinuity at $x=0$, at its instant of creation \cite{comment}.
Assuming that shock creation is uniformly distributed in time, one may write, for the probability density functional
to get configuration $u_0(x)$ at time $t=0$ (see appendix A),
\bea
Z_s &=&  \gamma \int_0^{\infty} \frac{d\eta}{\eta} \int_0^{\eta} dT  \int Du_s(x) \nonumber \\
&\times& P[u_s(x)] W[\eta,T;u_0(x),u_s(x)] \ , \ \label{z-shocks}
\eea
where $Du_s(x)P[u_s(x)]$ is the probability measure for the creation of the shock $u_s(x)$, conditioned to be in the
sample space of all shock creation events, and $W[\eta,T;u_0(x),u_s(x)]$ is a weighting functional.

Eq. (\ref{z-shocks}) is formally rigorous, but it is of hard practical implementation, due to the difficulty in getting
information on the functionals $P[u_s(x)]$ and $W[\eta,T;u_0(x),u_s(x)]$. Phenomenological arguments, however, can be helpful 
in order to replace (\ref{z-shocks}) by more tractable expressions. 

Shocks are expected to have (i) mean interdistances of the order of the integral scale $L$ and (ii) lifetimes 
of the order of $T = L/U$, where $U$ is an estimate of the shock velocity jump. The prototypical Burgers 
shock is the stationary configuration
\be
u_s(x;U) = - U \tanh \left ( \frac{U}{2 \nu} x \right ) \ . \ \label{st_shock}
\ee
Even though (\ref{st_shock}) is a solution of the forceless Burgers equation, it can be used as a
local approximation to general viscous shocks around the position of velocity discontinuity. Suppose that 
at time $-T$ a configuration similar to (\ref{st_shock}) is created, and let $g(U)$ be the probability density 
that it has amplitude $U$. Due to property (ii) above, this shock is not going to be observed at time $t=0$ if 
$T \gtrsim L/U$. The contribution to (\ref{z-shocks}) provived by shocks with the local profile (\ref{st_shock}) 
is, then, estimated as
\bea
Z_s &=& \gamma \int_0^\infty dU g(U) \frac{U}{L} \int_0^{L/U} dT {\cal{N}} \int D \hat u D u \nonumber \\
&\times& \delta[ \int dx \hat u(x,t)] \delta [ u(0,t) ]
 \exp (i S )  \ . \ \label{msr-phi-psib}
\eea
The velocity field $u(x,t)$ in (\ref{msr-phi-psib}) satisfies the boundary condition
\be
u(x,-T)= u_s(x;U) \ . \ \label{bound-c}
\ee

We will work with the phenomenologically simplified result (\ref{msr-phi-psib}) in place of (\ref{z-shocks}). 
In doing so, we conjecture that the asymptotic statistical properties of negative velocity-differences are not affected by more 
detailed choices of shock parametrization.

An interesting way to address the computation of (\ref{msr-phi-psib}) is to perform the corresponding path-integration over an appropriate
subset of the functional space, which would consist of dominating configurations. That's precisely the purpose of the saddle-point 
method, applied in Refs. \cite{falko_etal,gura-migdal} to the MSR turbulence context as a way to cope with the intermittency phenomenon.

Saddle-point configurations, dubbed instantons, are associated with stationary values of the action. 
Taking functional derivatives of the MSR action (\ref{msr-action}) with respect to the integration fields, we get
\bea
\partial_t u+ u \partial_x u - \nu \partial^2_x u  = -i\int dx'\hat u(x',t) D(|x-x'|)  \nonumber \\
\label{sp1}
\eea
and
\be
\partial_t \hat u+ u \partial_x \hat u + \nu \partial^2_x \hat u = 0  \ . \  \label{sp2}
\ee
Also, when solving (\ref{sp1}) and (\ref{sp2}), we have to take into account 
the constraints \cite{comment2}
\bea
&&u(0,t) = 0 \ , \ \label{sp3} \\
&&\int dx \hat u(x,t)  = 0 \ . \ \label{sp4}
\eea
Instanton solutions of (\ref{sp1}) and (\ref{sp2}), which hold for $-T < t < 0$ and satisfy 
(\ref{bound-c}), (\ref{sp3}) and (\ref{sp4}) can be readily obtained:
\bea
&&u(x,t)  = u_s(x;U) \ , \ \nonumber \\
&&\hat u(x,t)=0 \ . \ \label{shock}
\eea
It is worth mentioning that the solution for $u(x,t)$ in (\ref{shock}) identifies Burgers shocks to instantons. 
Also, it is not difficult to find that the MSR action vanishes when evaluated for the fields given in 
(\ref{shock}). As it is the standard procedure in the saddle-point method, we expand the MSR action in a 
functional Taylor series around the instantons, retaining only quadratic fluctuations. We replace, as a result,
(\ref{msr-action}) by
\bea
&&S^\star = \int_{-T}^0 dt \int dx \hat u [ \partial_t u+ \partial_x(u_s u) - \nu \partial^2_x u ]
\nonumber \\
&& + \frac{i}{2} \int_{- T}^0 dt \int dx dx' \hat u(x,t) \hat u(x',t) D(|x-x'|) \ , \ \nonumber \\
\label{msr-action2b}
\eea
where the velocity boundary condition becomes, now, $u(x,-T) = 0$.

In order to compute the pdf of negative velocity-differences, we introduce the characteristic function
\bea
&&Z_s (\lambda) = \gamma \int_0^\infty dU g(U) \frac{U}{L} \int_0^{L/U} dT {\cal{N}} \int D \hat u D u \nonumber \\
&&\times \delta[ \int dx \hat u(x,t)] \delta [ u(0,t) ] \exp (i S^\star - i \lambda z )  \ , \ \label{cf}
\eea
where $z$ is the velocity-difference evaluated at $t=0$,
\be
z = -2U + u(\zeta,0) -u(-\zeta,0) \ . \
\ee
The characteristic function $Z_s (\lambda)$ can be exactly computed, in principle, since it is 
given in (\ref{cf}) by a quadratic field theory. To evaluate $Z_s (\lambda)$, the saddle-point method can be 
applied once again, this time in an exact way. The further saddle-points equations for $u(x,t)$ 
and $\hat u(x,t)$ are
\bea
&&\partial_t u+ \partial_x (uu_s) - \nu \partial^2_x u  = \nonumber \\
&&= -i\int dx'\hat u(x',t) D(|x-x'|)  \ , \  \label{sp1b} \\
&&\partial_t \hat u+ u_s \partial_x \hat u + \nu \partial^2_x \hat u = \nonumber \\
&&= \lambda [\delta(x+\zeta) - \delta(x-\zeta)] \delta(t) \ , \  \label{sp2b}
\eea
supplemented by (\ref{sp3}) and (\ref{sp4}). 

Observe that the viscosity term has the ``wrong" sign in equation (\ref{sp2b}). To avoid the unbounded
growing of $\hat u(x,t)$ for $t>0$, we impose, as prescribed in Refs. \cite{falko_etal,gura-migdal}, the 
boundary condition $\hat u(x,0^+)=0$. Integrating (\ref{sp2b}) over the time interval 
$[ - \epsilon, \epsilon]$, with $\epsilon \rightarrow 0$, we get the ``final condition"
\be
\hat u (x,0^-) = \lambda [ \delta(x-\zeta)-\delta(x+\zeta)]  \ . \ \label{boundc-hat_u}
\ee
Furthermore, we have the exact saddle-point result
(see appendix B)
\bea
&& \bar S^\star - \lambda \bar z = 2 \lambda U \nonumber \\
&& + \frac{i}{2} \int_{- T}^0 dt \int dx dx' \hat u(x,t) \hat u(x',t) D(|x-x'|) \ . \ \nonumber \\
\label{sp-action1}
\eea
It is interesting to note, due to (\ref{sp-action1}), that we do not have to worry in finding
the specific solution for $u(x,t)$. Eq. (\ref{sp2b}) is solved, in the vanishing viscosity limit, by
\be
\hat u(x,t) = \lambda [ \delta(x-x(t))-\delta(x+x(t))] \ , \  \label{sp2sols}
\ee
where $x(t) = \zeta - Ut$. Substituting (\ref{sp2sols}) into (\ref{sp-action1}), and taking 
$\zeta / L <<1$, we find
\bea
&& \bar S^\star - \lambda \bar z = \nonumber \\
&&= 2 \lambda U + i \frac{D_0 L}{2U}\lambda^2 \int_{-2UT/L}^0 dt [ 1 - e^{-t^2}]  \ . \ 
\label{sp-action2}
\eea
We get, from (\ref{cf})
\bea
&&Z_s(\lambda) =  \gamma \int_0^\infty dU g(U) \int_0^1 d \eta \nonumber \\
&& \times \exp \left ( 2i\lambda U - \frac{D_0 L c(\eta)}{2U} \lambda^2 \right ) \ , \  \label{cf2}
\eea
where
\be
c(\eta) = \int_{-2 \eta}^0 dt [1 -e^{-t^2}] \ . \
\ee

The negative velocity-difference pdf is computed from the Fourier transform of the characteristic
function, as
\bea
&&\rho(z) = \frac{1}{2 \pi} \int_{-\infty}^{\infty} d \lambda \exp(i \lambda z) Z_s( \lambda) \nonumber \\
&&= \gamma \int_0^\infty dU g(U) \int_0^1 d \eta \sqrt{\frac{U}{2 \pi c(\eta) D_0 L}} \nonumber \\
&& \times \exp \left [ - \frac{U(z+2U)^2}{2 c(\eta)D_0 L} \right ] \ . \ \label{pdf}
\eea
Expression (\ref{pdf}) gives, for large negative $z$, the asymptotic pdf,
\be
\rho(z) = \frac{a}{|z|^3} + {\hbox{...}} \ , \
\ee
where the dots refer to subleading contributions, and
\be
a = \gamma L D_0 g(0) \int_0^1 d \eta c(\eta)  \simeq 0.36 \gamma L D_0 g(0) \ . \ \label{a_coef}
\ee 
The expression for the coefficient (\ref{a_coef}) is a testable prediction of the present theory. Alternative force-force 
correlation functions can be used to recompute (\ref{a_coef}) and compare it with the value to be found in further numerical 
simulations. It is clear that in the eulerian framework, the intermittency factor $\gamma$ has to be replaced by
\be
\gamma' = 2 n \zeta  \ , \
\ee
where $n$ is the number of shocks per unit length.

\section{Conclusions}
We have obtained, with the help of instanton techniques, the asymptotic form of the
pdf of large negative velocity differences in Burgers turbulence. The lagrangian picture 
of the Burgers flow was adapted to the MSR field theoretical framework, a procedure which 
proved to be an important technical improvement over the eulerian description. Lagrangian 
methods are, as a rule, welcome in the study of small-scale intermittency, since they cope 
in a natural way with the sweeping produced by large scale motions. In the case of Burgers 
flow, sweeping produces shock advection, making it difficult to find out the statistical 
properties of velocity difference fluctuations. 

The introduction of lagrangian coordinates was carried out under the hypothesis
of strong galilean invariance. We have found that the left tail pdf has the 
asymptotic form $\rho(z) = a/|z|^3$, which agrees with the conjecture put forward
in Ref. \cite{boldy-etal}, that this is so when strong galilean invariance holds.
We have obtained an explicit expression for the critical amplitude $a$, which
motivates the study of further numerical simulations of Burgers turbulence.

Arbitrary shocks of the Burgers forceless model are identified to instantons, and taken, 
in the path-integral formulation of the response functional, as the dominant configurations 
for the determination of the velocity difference fluctuations. We have bypassed the detailed 
classification of all of these Burgers shocks at their creation events, by noting that
relevant parameters of newborn shocks are their velocity jump, $U$, and extension, assumed to 
scale with the integral length $L$. Shocks are expected to have lifetimes of the order of $L/U$.
In view of the role of the dimensional parameters $U$ and $L$, we should regard the stationary 
Burgers shock (\ref{shock}) more as an illustration than as an essential ingredient in the 
formalism.

We emphasize that the instanton distribution $g(U)$ is not able to yield the left tail pdf of velocity 
differences on its own. Furthermore, the instanton contribution to the MSR action vanishes. The point is that
the fluctuating background couples with the shocks and by the usual instability mechanism discussed in 
section II, large negative velocity differences are generated in the flow. One may wonder if this process of 
intermittency generation is analogously found in the interaction between the background and coherent structures 
in Navier-Stokes turbulence.

\acknowledgements
This work has been partially supported by CNPq and FAPERJ.

\appendix

\section{Derivation of Eq. (\ref{z-shocks})}

Let us focus our attention on a flow which evolves under a particular realization of the stochastic force $f(x,t)$. We also suppose that a random ensemble of initial configurations is given in the remote past ($t \rightarrow - \infty$), so that at any instant of time the possible velocity configurations yield a statistically stationary ensemble.

We consider, now, the creation, at time $-T$, of a shock $u_s(x)$, localized at the origin of the locally comoving reference frame, which survives until time $t=0$. Let $T[u_s,f]$ be the maximum value of $T$. Since the probability for the creation of the shock $u_s(x)$ in a time interval $dT$ is also proportional to $dT$, we may write
\bea
Z_s &=& \gamma \int Du_s(x) P[u_s(x)] \nonumber \\
&\times& \langle \frac{1}{T[u_s,f]} \int_0^{T[u_s,f]} dT P[u_0,u_s;0,-T[u_s,f]] \rangle_f \ , \ \nonumber \\
\label{a1}
\eea
where $P[u_0,u_s;0,-T[u_s,f]]$ is the probability distribution associated with the transition from the shock configuration $u_s(x)$,
which was created at time $-T$, to the final configuration (at time $t=0$) $u_0(x)$. For the sake of clarity, we note that
\be
P[u_0,u_s;0,-T[u_s,f]] = \delta[u_0(x) - L(x;[u_s,T,f])] \ , \
\ee
where $L(x;[u_s,T,f])$ is the velocity configuration which evolves from the shock $u_s(x)$ after the time interval $T$.

Equation (\ref{a1}) can be rewritten as (\ref{z-shocks}). In fact,
\bea
Z_s &=& \gamma \int Du_s(x) P[u_s(x)]\int_0^\infty \frac{d\eta}{\eta} \int_0^{\eta} dT \nonumber \\
&\times& \langle \delta(\eta-T[u_s,f])P[u_0,u_s;0,-T] \rangle_f \nonumber \\
&=& \gamma \int_0^{\infty} \frac{d\eta}{\eta} \int_0^{\eta} dT  \int Du_s(x) \nonumber \\
&\times& P[u_s(x)] W[\eta,T;u_0(x),u_s(x)] \ , \ \label{a2}
\eea
where
\bea
&&W[\eta,T;u_0(x),u_s(x)] \nonumber \\
&&= \langle \delta(\eta-T[u_s,f])P[u_0,u_s;0,-T] \rangle_f \ . \ \label{a3}
\eea

Observe that if $T[u_s,f]$ does not depend on $f(x,t)$ (which may be a useful 
approximation), then 
\bea
&&W[\eta,T;u_0(x),u_s(x)] \nonumber \\
&& = \delta(\eta-T[u_s]) \langle P[u_0,u_s;0,-T] \rangle_f \nonumber \\
 \label{a4}
\eea
and, therefore,
\bea
Z_s &=& \gamma \int Du_s(x) P[u_s(x)] \frac{1}{T[u_s]} \int_0^{T[u_s]} dT \nonumber \\
&\times& \langle P[u_0,u_s;0,-T] \rangle_f  \ . \ \label{a5}
\eea
Above, the averaged probability $\langle P[u_0,u_s;0,-T] \rangle_f$ can be given as 
the MSR path-integral expression (\ref{msr}).

\section{Derivation of Eq. (\ref{sp-action1})}

There is some subtlety in the saddle-point evaluation of characteristic functionals like (\ref{cf}). 
Since we are considering in (\ref{cf}) the evolution up to time $t=0$, one could object that
the $\hat u(x,0^+)=0$ boundary condition sounds too loose. Actually, in order to apply the saddle-point 
method to (\ref{cf}), the time evolution is extended to $t\rightarrow \infty$. Saddle-point solutions are, 
then, such that $u(x, t \rightarrow \infty) = \hat u(x, t \rightarrow \infty) = 0$.  Note that the time
extension does not change the value of $Z_s(\lambda)$, once velocity configurations are integrated 
out at $t \rightarrow \infty$ in the path-integral (\ref{cf}).

Taking these remarks into account, we multiply both sides of Eq. (\ref{sp2b}) by $u(x,t)$ and integrate them 
over space and time. We find
\bea
&&\int dx \int_{-T}^\infty dt \hat u [\partial_t u+ u_s \partial_x u - \nu \partial^2_x u ] \nonumber \\
&&= \lambda [u(\zeta,0) - u(-\zeta,0)]  \ , \  \label{b1}
\eea
where we have used the boundary conditions $u(0,x)=0$, $\hat u(x,t>0)=0$ and $u(x,-T)=0$. Eq. (\ref{sp-action1}) follows straightforwardly from the substitution of (\ref{b1}) in the expression for $S^\star - \lambda z$, as it enters into (\ref{cf}).


\begin{references}
\bibitem{comment} New shocks can be created either from the smooth background, or from the merging of two moving shocks.
Conversely, the life of a shock ends when it just vanishes in the smooth background, or when it collides with 
another shock. These processes can be appreciated in an interesting movie produced by J. Bec, quoted in \cite{bec2}.
\bibitem{comment2} Since we find extrema of the action which satisfy (\ref{sp3}) and (\ref{sp4}), we do not use the method
of Lagrange multipliers. This simplification follows when one works with regularized delta functionals associated to the 
constraints, and removes the regulator after the saddle-point computation is performed.
\end{references}
\end{document}